\documentclass[a4paper,12pt]{amsart}
\usepackage{latexsym}
\usepackage[english]{babel}
\usepackage{graphicx}
\usepackage{psfig}

\begin{document}
\pagenumbering{roman}
\title[Region of the anomalous compression under Bondi-Hoyle accretion]
{Region of the anomalous compression under Bondi-Hoyle accretion}
\maketitle \fontsize{12pt}{\baselineskip}

\begin{center}
\textbf{Roman V. Shcherbakov}

\emph{Moscow Institute of Physics and Technology, Lebedev Physics
Institute}

{avalon@lpi.ru, shcher@gmail.com}

\end{center}

\abstract{We investigate the properties of an axisymmetric
non-magnetized gas flow without angular momentum on a small
compact object, in particular, on a Schwarzschild black hole in
the supersonic region near the object; the velocity of the object
itself is assumed to be low compared to the speed of sound at
infinity. First of all, we see that the streamlines intersect
(i.e., a caustic forms) on the symmetry axis at a certain distance
$r_x$ from the center on the front side if the pressure gradient
is neglected. The characteristic radial size of the region, in
which the streamlines emerging from the sonic surface at an angle
no larger than $\theta_0$ to the axis intersect, is $\Delta r=
r_x\theta^2_0/3.$ To refine the flow structure in this region, we
numerically compute the system in the adiabatic approximation
without ignoring the pressure. We estimate the parameters of the
inferred region with anomalously high matter temperature and
density accompanied by anomalously high energy release.}
\endabstract

\section{INTRODUCTION}

Bondi-Hoyle accretion is the fall of matter onto a moving compact
object. Although this problem was formulated in the mid-20th
century (Bondi and Hoyle 1944), many details of this process are
still incomprehensible. In particular, there is virtually no
detailed information about the properties of a flow in the
supersonic region near a gravitating center in the model of smooth
passage of the sonic surface (Bondi 1952). The first approximation
to spherically symmetric accretion was first considered to
calculate the accretion onto a nonrotating compact object moving
through a gas in the article (Beskin and Pidoprygora 1995). In
this case, the ratio of the object's velocity $v_\infty$ to the
speed of sound at an infinite distance from the body $c_\infty$
plays the role of a small parameter:
$$\varepsilon=\frac{v_\infty}{c_\infty}.$$

We consider only the first approximation with $\varepsilon<1$.
This inequality holds for certain astrophysical objects. Accretion
without shock formation is possible in this case. Since
spherically symmetric smooth transonic accretion was shown (Bondi
1952; Garlick 1979) to be stable, it has a physical meaning. It
can then be assumed that Bondi-Hoyle accretion without shock
formation is also stable at fairly small $\varepsilon$ and, hence,
also has physical meaning.

Recall the main properties of the smooth spherically symmetric
solution denoted by the superscript (0) and the first
approximation to it denoted by the superscript (1) (Bondi 1952;
Beskin and Pidoprygora 1995). The solution is sought in the
adiabatic approximation with a constant adiabatic index $\Gamma$.
All of the coefficients $k$ with different subscripts that appear
below were calculated by Beskin and Pidoprygora (1995) and depend
only on the adiabatic index $\Gamma$.

(1) The requirement of smoothness, i.e., the absence of shocks,
leads to an additional condition on the sonic surface that yields
$r_\ast=r^{(0)}_\ast+r^{(1)}_\ast$ for the radius of the sonic
surface in the nonrelativistic limit, where in polar
$(r,\theta,\phi)$ coordinates
$$r^{(0)}_\ast=\left(\frac{5-3\Gamma}{4}\right)\frac{G m}{c^2_\infty},
\qquad r^{(1)}_\ast=\varepsilon\cdot r^{(0)}_\ast
\left(\frac{\Gamma+1}{5-3\Gamma}\right)k_1(\Gamma)\cos\theta.$$

(2) The Grad-Shafranov equation defining the stream function
$\Phi(r,\theta)$ has a solution $\Phi=\Phi^{(0)}+\Phi^{(1)}$,
where
$$\Phi^{(0)}=\Phi_0(1-\cos\theta),\qquad \Phi^{(1)}=\Phi_0
\varepsilon g(r) \sin^2\theta.\quad \eqno(1)$$ Here $2\Phi_0$ is
the accretion rate, the stream function is defined as
$n\mathbf{v}=\mathbf{\nabla}\Phi\times \mathbf{e}_\phi/2\pi r \sin
\theta$, and $\mathbf{e}$ is a unit vector along the $\phi$ axis.

(3) The radial function $g(r)$ behaves asymptotically as
$g(r)=K(\Gamma)\left(r/r_\ast\right)^2$ far from the sonic
surface, which corresponds to a homogeneous flow. On the other
hand, in the supersonic region $r\ll r_{\ast}$
$$g(r)=k_{\rm in}(\Gamma)\left({r}/{r_{\ast}}\right)^{-1/2},\eqno(2)$$
 where $r_{\ast}$ is a radius of the sonic surface.

(4) We see that the perturbation  $\Phi^{(1)}$ becomes significant
in the supersonic region, and $\max_{\theta}(\Phi^{(1)}/\Phi_0)$
reaches unity at
$$r_x=r_\ast(\varepsilon k_{\rm in})^2.\eqno(3)$$
 So, the
perturbation theory does not hold there and we have to solve exact
equations.

In this paper, we describe the method used and the simplifications
that help to carry out calculations. We prove that these
simplifications can be introduced without any significant loss of
the accuracy and give basic formulas and results. Subsequently, we
discuss whether this phenomenon is observable.

\section{DESCRIPTION OF THE METHOD}

\bigskip

The essence of the described method is to directly calculate the
streamline, which allows the physically observable quantities to
be easily found. This method is also the most natural for deriving
clear equations.

Let us parameterize the flow line as ${\theta =\theta
({\theta_0},r) }$. The initial angle $\theta_0$ at some radius
$r_0$ and radius $r$ are independent variables. We choose $r_0$ in
the supersonic region where it makes sense to consider the first
approximation. The thermodynamic potentials are functions of the
same arguments.

First of all, note that this parametrization is unique. However,
to specify the flow, we must specify not only the trajectory, but
also the velocity along it $v[\theta(\theta_0,r),r]$. In this
case, two differential equations for the two functions $v$ and
$\theta$ can be derived. They should then be solved. These are the
energy equation and the equation of forces or angular momentums.

The problem is solved by assuming the absence of energy release at
the stellar boundary; i.e., the object is actually assumed to be a
black hole. There is no angular momentum along the $\theta = 0$
axis.

\section{SIMPLIFYING ASSUMPTIONS}

Let us introduce several simplifications none of which, as we will
see, affects significantly the result.

(1) In our calculations, we use the metric of flat space.

(2) The radial velocity of the matter $v_r$ is much higher than
the nonradial velocity $v_\theta$. Therefore, the tangential
components of the gradients for all parameters of the system are
much smaller than their radial components.

(3) To determine the radial velocity, we neglect the enthalpy of
the matter compared to the gravitational energy.

(4) The solution can be represented as a converging series in
$\theta_0$ in the form
$$\theta=\sum _{n=0}^{{\infty}}{{{{\theta
}_0}}^{2n+1}}{k_{2n+1}}(r)$$ on the front side of a compact object
near the symmetry axis ($\theta(0,r)=0$ on the axis).

\section{BASIC EQUATIONS}

Let us first solve the problem by disregarding the pressure in the
supersonic region. We go to the frame of reference, in which the
body is at rest and the gas flows on it. Let us write the angular
momentum conservation equation for the gas relatively to the
center of the object as
$$\frac{d}{dt}\left(r^2\frac{d
  \theta}{dt}\right)=0.\eqno(4)$$
Next, we can write the expression for the radial velocity as
$$v_\text{full} \approx |v_r|=-\frac{dr}{dt}\approx{\sqrt{\frac{{r_{\rm
g}}}{r}}},\eqno(5)$$ where $r_{\rm g} = 2GM/c^2$ is the
gravitational radius of an object of mass $M$. Eliminating $dt$
from equations (4,5), we immediately determine the streamline as
$$\frac{\partial}{\partial r}\left(r^{3/2}\frac{\partial \theta}{\partial r}\right)=0.$$
The solution of this equation that satisfies the natural initial
condition $\theta(\theta_0,r_0)=\theta_0$ is $$\quad
\theta=\theta_0 -
C(\theta_0)\left[\left(\frac{r_0}{r}\right)^{1/2}-1\right].\eqno
(6)$$

We derive the form of the function $C(\theta_0)$ from (2), which
yields the same solution for the streamline, but has a limited
validity range as the first approximation. As a result, we obtain

$$ \theta(\theta_0,r)=\theta_0 -\varepsilon
k_{\rm in}\sqrt{r_\ast}
\left(\frac{1}{\sqrt{r}}-\frac{1}{\sqrt{r_0}}\right) \sin\theta_0
.$$

This expression has a nontrivial structure. It defines a caustic
near the distance $R_{\rm ball}$ from the center,
$$R_{\rm ball}=r_\ast(\varepsilon k_{\rm in})^2.$$
For physically meaningful $\Gamma$ ($\Gamma > 1.25$), this caustic
is located on the front side of the object, although the saddle
point with a zero gas velocity is always on the rear side. The
characteristic radial size of the region, in which the streamlines
emerging from the sonic surface at an angle no larger than
$\theta_0$ to the axis, intersect is $\Delta r= r_x\theta^2_0/3$.
Naturally, the gas pressure shouldn't be ignored in a region of
significant compression; the trajectories will not intersect if
the pressure is included.We can qualitatively predict that, in
this case, a certain region of strong compression will be closer
to the object than $R_{\rm ball}$. Next, let us perform a
calculation without neglecting the pressure. Let us set up a
continuity equation for the gas flow. The area of the
cross-sectional element of the flow in a direction perpendicular
to the radial vector is
$$  \delta  S= 2\pi {r^2}\sin\theta  \delta  \theta .\eqno(7)$$
The continuity equation itself from which we find the number
density $n$ can then be written as
$$v_r n\delta S ={\rm const}.\eqno(8)$$
Using (5) and (7), we obtain from (8)
 $$  {r^{3/2}}n\sin \theta
\delta \theta ={{{r_0}}^{3/2}}n_0\sin {\theta_0}\delta
{\theta_0},$$ where $\theta$ is the function of $\theta_0$ at
$r_0.$ We then get
$$
 \frac{\delta
 \theta }{\delta  {\theta_0}}={{\partial }_{{\theta_0}}}\theta ({\theta_0},r).
 $$

Hence, the number density can be written as
$$n=n_0\left(\frac{r_0}{r}\right)^{3/2}\frac{\sin\theta_0}{\sin\theta}\frac{1}{\partial_{\theta_0}\theta}.\eqno(9)$$
Below, for simplicity, we consider an adiabatic process on an
ideal gas, $p=k(s) n^{\Gamma}$, where the entropy $s$ is constant
in the entire space. The pressure, the temperature, and the speed
of sound can then be expressed using (8).

We can now calculate the curvature of the streamline under
pressure. The point probe mass is affected by the gravity force
and the force proportional to $\mathbf{\nabla} p$. We can derive
an equation for gas angular momentum deviation relatively the body
center; the gravity torque is zero indeed. The angular momentum is
the vector perpendicular to the plane passing through the symmetry
axis. Only the tangential part of $\mathbf{\nabla} p$ will appear
in what follows.

The moment equation is  $$ \frac{d}{{dt}}{\Delta L}=-{\Delta K},$$
where $L$ is the angular momentum, and $K$ is the moment of force.
The minus corresponds to the repulsion of streamlines.

 The force acting upon the small element is ${\vspace{0.666667ex}}{\Delta F}=l {\Delta r} {\Delta
p} $, where $l$ is the length of the circumference described in
the plane perpendicular to the $\theta = 0$ axis. The
corresponding moment of force is $ {\Delta K}=l r {\Delta r}
{\Delta p},  $ and the mass of the element is $ {\Delta m}={m_{\rm
p}}n l r {\Delta \theta } {\Delta r}$, where $m_{\rm p}$ is the
mass of a single particle. The angular momentum of this probe mass
is $\Delta L = m_{\rm p} n l r \Delta \theta \Delta r \cdot
r^2{d\theta/dt}.$ Finally, we find that
$$\frac{d}{dt} \left(r^2 \frac{d\theta}{dt}\right)=-\frac{\Delta
p}{\Delta \theta m_{\rm p} n}. $$ At $r=\rm const$ we obtain
$p=p(\theta_0)$ and $\theta=\theta(\theta_0)$; therefore $\Delta
p/\Delta \theta=\partial_{\theta_0}p(\theta_0,r) /
\partial_{\theta_0}{\theta(\theta_0,r)}$.
We replace the differentiation with respect to $t$ by the
differentiation with respect to $r$ using (5):
$$ {{\partial }_r}\Big({r^{\frac{3}{2}}}{{\partial }_r}\theta
({\theta 0},r)\Big)=-\frac{{{\partial }_{{{\theta }_0}}}p({{\theta
}_0},r){\sqrt{r}}}{{m_p}n{{\partial }_{{{\theta }_0}}}\theta
({{\theta }_0},r){r_g}}\eqno(10)$$ Substitution in the number
density in form (9) yields
$$\displaystyle{ {{\partial }_r}\left[{r^{3/2}}{{\partial }_r}\theta
({{\theta }_0},r)\right]=-\sqrt{r_0}\frac{{{{c_0}}^2}}{{r_{\rm
g}}}{{\left(\frac{r}{{r_0}}\right)}^{\frac32(\frac43-\Gamma)
}}\frac{{{\left[\frac{\sin \theta ({{\theta }_0},r)}{\sin {{\theta
}_0}}{{\partial }_{{{\theta }_0}}}\theta ({{\theta
}_0},r)\right]}^{2-\Gamma }}}{({{\partial }_{{{\theta }_0}}}\theta
({{\theta }_0},r))}{{\partial}_{{{\theta }_0}}}\bigg(\frac{\sin
{{\theta }_0}}{\sin \theta ({{\theta }_0},r)}\frac{1}{{{\partial
}_{{{\theta }_0}}}\theta ({{\theta }_0},r)}\bigg) }.\eqno(11)$$
This expression can be rewritten in convenient dimensionless
coordinates. Let us denote  $$x=\frac{c^2_\ast r}{4 r_{\rm g}}.$$
So $x_\ast=1$, and $x_0=(r_0/r_\ast)$. As a result, equation (11)
takes the form

$$\displaystyle{ {{\partial }_x}\left[{x^{3/2}}{{\partial
}_x}\theta ({{\theta
}_0},x)\right]=-4{{\left(\frac{x}{{x_0}}\right)}^{\frac{3}{2}(\frac43-\Gamma)
}}\frac{{{\left[\frac{\sin \theta ({{\theta }_0},x)}{\sin {{\theta
}_0}}{{\partial }_{{{\theta }_0}}}\theta ({{\theta
}_0},x)\right]}^{2-\Gamma }}}{({{\partial }_{{{\theta }_0}}}\theta
({{\theta }_0},x))}{{\partial }_{{{\theta }_0}}}\bigg(\frac{\sin
{{\theta }_0}}{\sin \theta ({{\theta }_0},x)}\frac{1}{{{\partial
}_{{{\theta }_0}}}\theta ({{\theta }_0},x)}\bigg) }.\eqno(11')$$


We have obtained a second order partial differential equation of
the hyperbolic type that is linear in second partial derivatives.
Let us transform it to a system of ordinary differential
equations. For this purpose, we use assumption (4) and substitute
$\theta$ with a finite sum containing the terms with the powers of
$\theta$ up to $2N+1$ $${ \theta ({{\theta}_0},r)=\sum
_{n=0}^{{N}}{{{{\theta }_0}}^{2n+1}}{k_{2n+1}}(r) }.\eqno(12)$$ We
know the function $\theta(\theta_0,r)$ (10) at the radius $r_0$
which defines initial conditions for (18). Then
$$k_1(r_0)=1,\quad
k_{2n+1}(r_0)=0,\quad n=1,2,...,N .\eqno(13)$$
$${{k'}_{2n+1}}({r_0})={{(-1)}^n}\frac{\varepsilon k_{\rm in}}
{2(2n+1)!{r_0}}\sqrt{\frac{r_\ast}{r_0}},\quad n=0,1,...,N
.\eqno(14)$$ The Cauchy problem is Eq. (11) with the initial
conditions (13) and (14). After substituting (12), we obtain a
system for $N +1$ functions of the radius. The absence of terms
with even degrees can be explained by the zero initial conditions
for them, while the equation to be solved is homogeneous.

The seeming complexity of the method of solution described above
can be easily explained when it is considered that we must
separate out a region of a very small size in both è $\theta\ll1$
 and $r$ when solving the second-order partial differential
equation. Qualitatively, the behavior of the trajectories can be
described as follows. A plane uniform flow at infinity transforms
into an almost spherically symmetric flow near the sonic surface.
Subsequently, depending on the adiabatic index, three cases are
possible (Beskin and Pidoprygora 1995):

 1). For $\Gamma\rightarrow 5/3$ the Mach number M, which
defines the ratio of the gravitational forces to the pressure
force, does not tend to infinity, but retains a value of the order
of unity as r approaches zero. It can be no significant additional
compression compared to that in the spherically symmetric case.

2). For $\Gamma<1.25 $ the parameter $k_{\rm in}(\Gamma)$ is
negative with positive $K(\Gamma)$, therefore, the streamlines
converge on the rear side (this can be seen from (1)).

3). For $\Gamma>1.25 $  the sign of the coefficient $k_{\rm in}$
is the opposite; the streamlines converge on the front side of the
object. After the passage of the region of maximum streamline
convergence near the symmetry axis, the pressure pushes them
apart.

Thus, our region is described by the following quantities: the
radius $r_x$, at which there is maximum additional compression on
the axis compared to that in Bondi accretion; the dimensionless
radius $x_x$; the radial size of the region $\Delta r $ at the
boundary of which the additional compression decreases by a factor
of 2; the caustic radius at the same parameters, but without
including the pressure $R_{\rm ball}$; the minimum achievable
ratio $\theta/\theta_0$ in this region $k_x$; the minimum Mach
number $M_x$ in the vicinity of $r_x$.

\bigskip
\bigskip
\bigskip
\bigskip\bigskip\bigskip\bigskip\bigskip

$\Gamma=4/3\qquad\qquad\qquad\qquad\qquad\qquad\qquad\qquad\qquad\quad\Gamma=7/5$

\begin{tabular}{|l|l|l|c|l|c|}
  \hline
  $\varepsilon $ & $x_x$ & $k_x$  &  $\frac{R_{\rm ball}}{r_x}$& $M_x$&$\frac{r_x}{r_{\rm g}}$ \\
    \hline
 1 & $2.5\cdot 10^{-4}$ &0.036  &2.5 &2.5&$7.8\cdot10^2$ \\
0.64 & $1.1\cdot 10^{-4}$ &0.029 &2.3 &2.9&$3.6\cdot10^2$  \\
0.4 &$4.8\cdot10^{-5}$ &0.023 & 2.1&3.4&$1.5\cdot10^2$ \\
0.24&$1.9\cdot10^{-5}$ &0.017&  1.9&4& 59\\
0.16&$8.9\cdot10^{-6}$&0.014&1.8&4.5&28\\
0.10&$3.5\cdot10^{-6}$&0.011&1.8&5.3&11\\
 0.06&$1.3\cdot10^{-6}$&0.008&1.7&6.1&4\\
  \hline
\end{tabular}$\quad\quad$
\begin{tabular}{|l|l|l|c|l|c|}
  \hline
  $\varepsilon $ & $x_x$ &$k_x$ & $\frac{R_{\rm ball}}{r_x}$ &$M_x$&$\frac{r_x}{r_{\rm g}}$  \\
    \hline
1 &$1.6\cdot 10^{-4}$ &0.068 &4.1 &2.0&$4\cdot10^2$  \\
0.58 &$5.8\cdot 10^{-5}$ &0.054  &3.9 &2.2&$1.4\cdot10^2$ \\
0.38 &$2.7\cdot 10^{-5}$ &0.044 & 3.7&2.3& 67\\
0.23&$1.1\cdot 10^{-5}$ &0.036& 3.3&2.6&  27\\
0.15&$5.2\cdot 10^{-6}$ &0.031&3.1&2.8&13\\
0.10&$2.3\cdot 10^{-6}$ &0.026&2.9&3.1&6\\
0.06&$8\cdot 10^{-7}$ &0.021&2.8&3.5&2\\
  \hline
\end{tabular}

\bigskip
Let us make a calculation for two cases: $\Gamma = 4/3$, $c_\infty
= 0.0002$ and $\Gamma = 7/5$, $c_\infty = 0.00001$. The first case
is nonphysical. It has only a theoretical significance as an
extension of classical astrophysical problems. In turn, the sonic
surface with the parameters corresponding to the second case can
be passed in dense clouds of molecular hydrogen. The
characteristic radial size of the region is $\Delta r \approx
r_x/5$. The solution is stable for $r>r_x$; i.e., the result does
not change with increasing $N$. Further out, the algorithm is
unstable; therefore, the behavior of the solution cannot be
determined at small $r\ll r_x$.

Let $k(r) = k_1(r)$, accordingly, $k_x = k(r_x)$. Let us estimate
the physical parameters of the system on the axis. First, the
number density can be written as
$$n(0,r)=n_0\left(\frac{r_0}{r}\right)^{3/2}k(r)^{-2},$$
which yields
$$ M=\frac{v}{{c_s}}\sim \frac{1}{{\sqrt{T}}{\sqrt{r}}}\sim
\frac{{n^{(1-\Gamma )/2}}}{{\sqrt{r}}}\sim  {r^{(3\Gamma
-5)/4}}{k^{\Gamma -1}} $$ for the Mach number. Thus
$M(0,r)={{\left(r_{kr}/{r}\right)}^{(5-3\Gamma
)/4}}{{k(r)}^{\Gamma -1}}\quad $ and
 $${M_x}\approx
\left(\frac{R_{\rm ball}}{r_x}\right)^{(5-3\Gamma
)/4}{{(\varepsilon {k_{{\rm in}}})}^{(3\Gamma
-5)/2}}{{k_x}^{\Gamma -1}}.$$

For $\Gamma=4/3$ we have $k_{\rm in}=0.025$, and for $\Gamma=7/5$
we have $k_{\rm in}=0.026$ (Beskin and Pidoprygora 1995). In fact,
the small parameter for $\Phi^{(1)}$ is $\varepsilon k_{\rm in}$
rather than $\varepsilon$, which has an order of $0.1\div 1$.

 \begin{figure}[h]
\includegraphics[height=10cm, width=16.6cm]{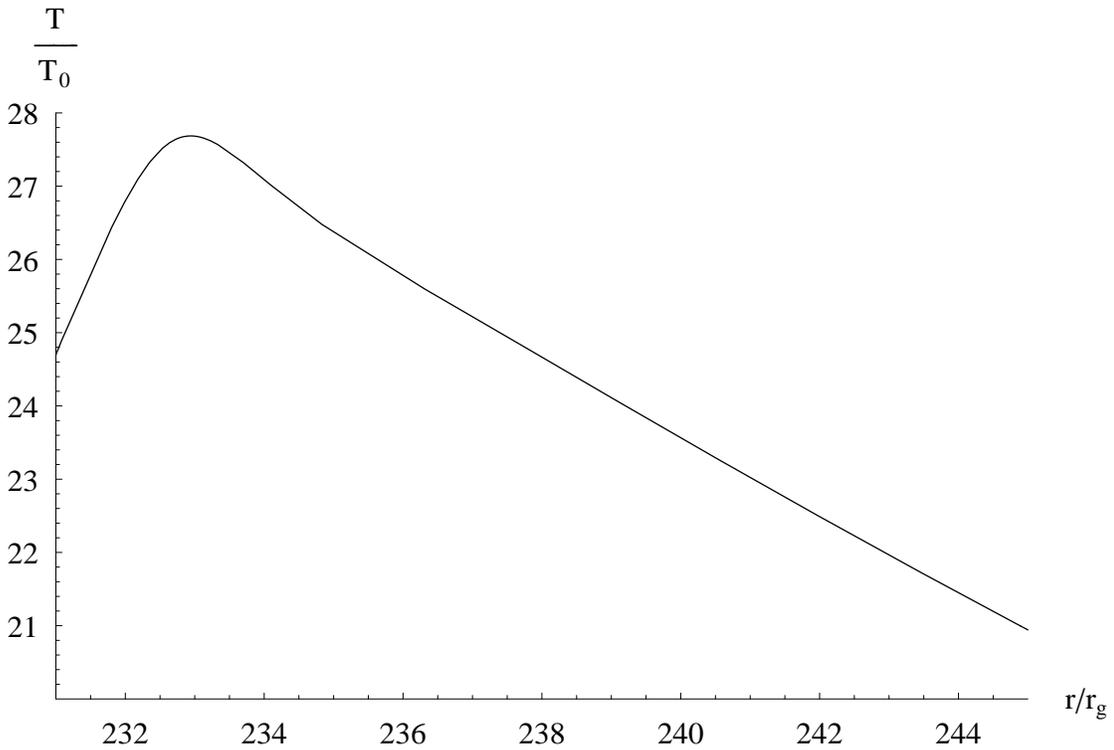}
 \caption{Temperature ratio along the $\theta = 0$ axis versus
 $r/r_g$
near the singularity for the parameters $\Gamma = 7/5$,
$\varepsilon = 0.03$, $c_\infty = 0.00001$; $T_0$ corresponds to
Bondi (1952) accretion.} \label{fig:fig. 1}
\end{figure}

Figure 1 shows the dependence of the ratio of the temperatures in
our case and in the spherically symmetric case on the
dimensionless distance to a compact object calculated for the
above parameters.

Figure 2 shows the corresponding bivariate dependence of the
temperature on the dimensionless distance and the angle
$\theta_0$. The transformation to the $(r/r_{\rm g}, \theta)$
plane near $r_x$ is to be made by a factor of $1/k_x$ compression
towards the symmetry axis of the system. The light spots in the
lower corners of the figure have no physical meaning, but are
determined by the divergence of the algorithm. Using Fig. 1, we
can associate the brightness in Fig. 2 with temperature.

 \begin{figure}[h]
\includegraphics{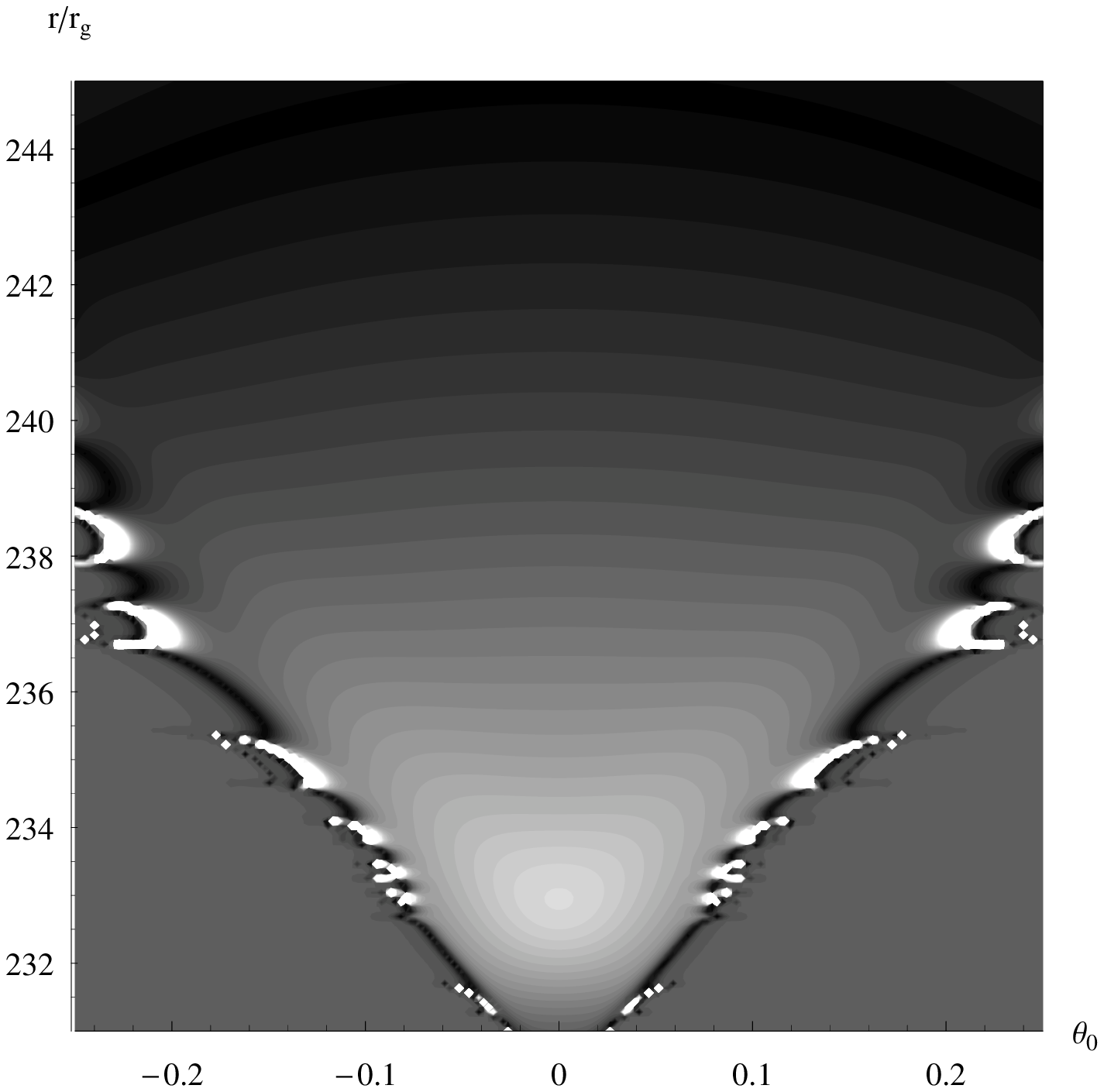}
 \caption{Temperature versus coordinates in the plane of variables $(r/r_{\rm g}, \theta_0)$ for the parameters
$ \Gamma = 7/5$, $\varepsilon = 0.03$, $c_\infty = 0.00001$. The
light area at the center has the highest temperature.}

 \label{fig:fig. 2}
\end{figure}

\section{VALIDITY OF THE ASSUMPTIONS}

To solve the problem, we made many assumptions. Let us
substantiate them. For our estimates, we set $c_\infty=0.0002$.

(1) In most cases, the needed region is located far from the event
horizon of the object, $r \gg r_g$, which allows us to consider
the metric flat.

(2) The inclination of the streamline á to the radial vector is
near $r_x$. The solution for $n(r)$ yields $\Delta
r=5\cdot10^5\text{cm}, r_x=10^7\text{cm},$ and $k_x=0.03$. The
region is compressed almost uniformly for $\theta_0<0.3.$ whence,
we can estimate $\alpha$ as $\alpha \approx\theta_0 k_x r_x/\Delta
r\approx 0.2$, and this is the maximum possible value for $\alpha$
within our range of parameters.

(3) To substantiate the possibility of ignoring the enthalpy of
matter compared to the gravitational energy, we calculated the
Mach number $M$. The ratio of the enthalpy to the kinetic energy
for a non-relativistic gas is $2
c_s^2/v^2(\Gamma-1)=2/M^2(\Gamma-1)$, reaching unity at $r_x$.
However, this is true in a small region. This effect can be easily
estimated by solving the equations for the effectively increased
coefficient of the term responsible for the pressure. When the
coefficients increases by a factor of $1.5$, $r_x$ decreases by
approximately the same factor $1.5$, while $k_x$ increases by no
more than $10\%$.

4. The smallness of ${\theta_0}^{25}k_{25}(r_x)$ compared to
$\theta_0 k_1(r_x)$ corresponds to $\theta_0<0.3$ at $r=r_x$,which
corresponds to an area fraction of $0.036$ of the initial one at
$r_0$. This implies that much $(> 1/30)$ of the flow is subjected
to strong compression. At a large radius, i.e., larger than $r_x$,
each succeeding term $\theta^{2n+1}r_{2n+1}$ is much smaller than
the preceding term, while this is not the case for $r < r_x$. This
is most likely because the convergence is lost for $r < r_x$ in
the method of solution used.

 5. It is necessary to check as well that the results don't depend on the $r_0$. Let us take
$\varepsilon=0.58$ and $\Gamma=7/5$: when $r_0$ doubles, $k_x$
decreases by $10\%$ and $r_x$ decreases by $20\%$. The difference
stems from the fact that the flow become slightly compressed,
while moving from the sonic surface to $r_0$ , and, most
importantly, its direction slightly changes. For smaller
$\varepsilon$, the $r_0$ dependence of the parameters of the
region is even weaker.

Note that the satisfactory accuracy of the results cannot be
improved significantly by abandoning the assumptions made in favor
of more accurate calculations. First, we used the first
approximation for our calculations with a finite $\varepsilon$,
and, second, we used its form at small radii where the firsts
approximation grows compared to the zeroth solution. We cannot
specify any initial conditions near the sonic surface, since the
behavior of formula (5) is asymptotic for $r\rightarrow0$.

To make progress in this field requires numerically solving an
exact system of two nonlinear partial differential equations, one
first-order equation and one second-order equation. In this case,
we must reveal features that are many orders of magnitude smaller
than the outer radius of integration or even the sonic surface,
which can be achieved only by using adaptive methods with a
variable step. In addition, we must numerically calculate the
parameters of the separatrix surface (the boundary of the causally
connected region) at which its smooth passage is realized, solve
the time-dependent problem of establishment of this regime, show
that it is stable, and find the percentage of cases, in which the
accretion without shocks is possible. Such calculations are far
beyond the scope of this paper.

\bigskip
\section{PHYSICAL AND OBSERVATIONAL SIGNIFICANCE OF THE RESULT}

Noteworthy among the features of the axisymmetric flow on a
compact object observed so far is the tail convergence of
streamlines behind a moving object during accretion in a region of
cold dense dust, where star formation takes place.

The effect discussed in this paper has not yet been observed. The
parameter $\Gamma = 7/5$ has a very narrow validity range. For
example, it is realized at the initial stage of accretion onto a
compact object from dense clouds of molecular hydrogen before its
dissociation (the equations with a constant $\GammaÃ$ cannot be
used at all when the dissociation occurs).

Let us consider in more detail the accretion of a hot gas from H
II regions with $\Gamma = 5/3$ far from the massive object. The
value of $\Gamma$ is the same on the sonic surface, since the gas
density and temperature at infinity and on the sonic surface
differ by a small factor (Bondi 1952). In this case, the angular
momentum of the gas relative to the stellar center in a stable
regime is very low. Although the electrons become relativistic and
the adiabatic index decreases sharply $\Gamma = 13/9$ starting
from a certain distance (Shapiro and Teukolsky 1985), no region of
anomalous compression is formed in this case.

Nevertheless, let us calculate the energy release in the region of
streamline convergence for nonphysical constant values of $\Gamma
< 5/3$. The calculation will yield an upper limit for actual
astrophysical objects. For our estimation, we assume that all of
the observed radiation in the derived temperature range is only
thermal bremsstrahlung (Shapiro 1973). Its intensity per unit
volume is defined by the non-relativistic formula $\Lambda =
cn^2T^{1/2}$ with the coefficient $c$ that depends weakly on the
number density $n$ and temperature $T$. The intensity of the
radiation for spherically symmetric accretion is then
$$L_0=\int\Lambda d^3r=\frac{8\pi}{3\Gamma-1} c n_x^{\Gamma+1} \left(\frac{r_x}{r_g}\right)^{3/2(\Gamma+1)}{r_g}^3.$$
The characteristic radial size of the region in our case is
$\Delta r \approx r_x/5$, and the tangential size is
$r_\perp\approx r_x \theta_x k_x$, where $\theta_x$ defines the
maximum angle at radius $r_0$, starting from which flow lines
contracts in $1/k_x$. The energy release $L_x$ in our region is
$$L_x\approx\Lambda_x\Delta r\pi r_\perp^2=C
n_x^{\Gamma+1}\pi\theta_x^2 k_x^{-2\Gamma}r_x^3.$$ For our
estimation, we set $\theta_x = 0.5$. The ratio of the released
energies can then be written as
$$\frac{L_x}{L_0}=\frac{3\Gamma-1}{160}k_x^{-2\Gamma}\left(\frac{r_x}{r_g}\right)^{3/2(1-\Gamma)}.\eqno(14)$$
It follows from (14) and from the tables that the ratio of the
released energies increases as the radius, at which the
streamlines are focused, decreases. Let us calculate it for the
minimum possible $\varepsilon$ when the focusing still takes
place.We take the lower rows from the tables of parameters for the
region of anomalous compression for $\Gamma = 4/3$ and $\Gamma =
7/5$. We find from (14) that $L_x/L_0$ reaches $100$; i.e., when
the focusing is present, the Bondi-Hoyle accretion efficiency can
increase up to a factor of $10^2$. For physically meaningful value
$\Gamma = 7/5$, the calculated ratio (14) actually specifies only
an upper limit for $L_x/L_0$, since the gas dissociates rapidly
and is subsequently ionized as it heats up, which ensures a
transition to the regime with $\Gamma = 5/3$ and is accompanied by
a significant decrease in $L_x$. However, the proper allowance for
the change in adiabatic index in the axisymmetric case is beyond
the scope of this paper. Although the velocity dispersion of
compact objects is much higher than the speed of sound in the
surrounding cloud of gas and dust, the described regime with
$\varepsilon < 1$ can take place in certain cases at a low
relative velocity of the gas cloud and the compact object. Such a
change in energy release and accretion spectrum must be taken into
account when analyzing experimental data to estimate the
parameters of the black hole itself moving in regions of gas
compression and its surrounding medium.

\section{Acknowledgements}
I wish to thank Prof. V.S. Beskin, my scientific adviser, for
fruitful discussions. This work was supported in part by the
Russian Foundation for Basic Research (project no. 05-02-17700).
\section{References}
1. V. S . Beskin and Yu. N. Pidoprygora, Zh. ´ Eksp. Teor. Fiz.
107, 1025 (1995) [JETP 80, 575 (1995)].

2. H. Bondi and F.Hoyle, Mon. Not. R. Astron. Soc. 104, 273
(1944).

3. H. Bondi, Mon. Not. R. Astron. Soc. 112, 195 (1952).

4. A. R. Garlick, Astron. Astrophys. 73, 171 (1979).

5. S. L. Shapiro, Astrophys. J. 180, 531 (1973).

6. S. L. Shapiro and S. A. Teukolsky, \textit{Black Holes,White
Dwarfs, and  Stars: the Physics of Compact Objects } (Wiley, New
York, 1983; Mir,Moscow, 1985).
\end{document}